\documentstyle[twoside]{article}

\catcode`\@=11
\long\def\@makefntext#1{ 
\protect\noindent \hbox to 3.2pt {\hskip-.9pt
$^{{\eightrm\@thefnmark}}$\hfil}#1\hfill} 

\def\thefootnote{\fnsymbol{footnote}}
 \def\@makefnmark{\hbox to 0pt{$^{\@thefnmark}$\hss}}  

\def\ps@myheadings{\let\@mkboth\@gobbletwo
\def\@oddhead{\hbox{} 
\rightmark\hfil\eightrm\thepage}
\def\@oddfoot{}\def\@evenhead{\eightrm\thepage\hfil 
\leftmark\hbox{}}\def\@evenfoot{}
\def\sectionmark##1{}\def\subsectionmark##1{}}

\oddsidemargin=\evensidemargin
\renewcommand{\thefootnote}{\fnsymbol{footnote}}

\newcounter{sectionc}\newcounter{subsectionc}\newcounter{subsubsectionc}
\renewcommand{\section}[1] {\vspace{12pt}\addtocounter{sectionc}{1}
\setcounter{subsectionc}{0}\setcounter{subsubsectionc}{0}\noindent
	{\tenbf\thesectionc. #1}\par\vspace{5pt}}
\renewcommand{\subsection}[1] {\vspace{12pt}\addtocounter{subsectionc}{1}
	\setcounter{subsubsectionc}{0}\noindent
	{\bf\thesectionc.\thesubsectionc. {\kern1pt \bfit #1}}\par\vspace{5pt}}
\renewcommand{\subsubsection}[1] {\vspace{12pt}\addtocounter{subsubsectionc}{1}
	\noindent{\tenrm\thesectionc.\thesubsectionc.\thesubsubsectionc.
	{\kern1pt \tenit #1}}\par\vspace{5pt}}
\newcommand{\nonumsection}[1] {\vspace{12pt}\noindent{\tenbf #1}
	\par\vspace{5pt}}

\newcounter{appendixc}
\newcounter{subappendixc}[appendixc]
\newcounter{subsubappendixc}[subappendixc]
\renewcommand{\thesubappendixc}{\Alph{appendixc}.\arabic{subappendixc}}
\renewcommand{\thesubsubappendixc}
	{\Alph{appendixc}.\arabic{subappendixc}.\arabic{subsubappendixc}}

\renewcommand{\appendix}[1] {\vspace{12pt}
        \refstepcounter{appendixc}
        \setcounter{figure}{0}
        \setcounter{table}{0}
        \setcounter{lemma}{0}
        \setcounter{theorem}{0}
        \setcounter{corollary}{0}
        \setcounter{definition}{0}
        \setcounter{equation}{0}
        \renewcommand{\thefigure}{\Alph{appendixc}.\arabic{figure}}
        \renewcommand{\thetable}{\Alph{appendixc}.\arabic{table}}
        \renewcommand{\theappendixc}{\Alph{appendixc}}
        \renewcommand{\thelemma}{\Alph{appendixc}.\arabic{lemma}}
        \renewcommand{\thetheorem}{\Alph{appendixc}.\arabic{theorem}}
        \renewcommand{\thedefinition}{\Alph{appendixc}.\arabic{definition}}
        \renewcommand{\thecorollary}{\Alph{appendixc}.\arabic{corollary}}
        \renewcommand{\theequation}{\Alph{appendixc}.\arabic{equation}}
        \noindent{\tenbf Appendix \theappendixc #1}\par\vspace{5pt}}
\newcommand{\subappendix}[1] {\vspace{12pt}
        \refstepcounter{subappendixc}
        \noindent{\bf Appendix \thesubappendixc. {\kern1pt \bfit #1}}
	\par\vspace{5pt}}
\newcommand{\subsubappendix}[1] {\vspace{12pt}
        \refstepcounter{subsubappendixc}
        \noindent{\rm Appendix \thesubsubappendixc. {\kern1pt \tenit #1}}
	\par\vspace{5pt}}

\newcommand{\textlineskip}{\baselineskip=13pt}
\newcommand{\smalllineskip}{\baselineskip=10pt}

\def\eightcirc{
\begin{picture}(0,0)
\put(4.4,1.8){\circle{6.5}}
\end{picture}}
\def\eightcopyright{\eightcirc\kern2.7pt\hbox{\eightrm c}}

\newcommand{\copyrightheading}[1]
	{\vspace*{-2.5cm}\smalllineskip{\flushleft
	{\tenrm USITP-#1}\\
	{\tenrm hep-th/9308041}\\
	 }}

\def\abstracts#1#2#3{{
	\centering{\begin{minipage}{4.5in}\baselineskip=10pt\eightrm
	\centerline{ABSTRACT}
	\parindent=0pt #1\par
	\parindent=15pt #2\par
	\parindent=15pt #3
	\end{minipage} }\par}}

\newcommand{\bibit}{\nineit}

\renewenvironment{thebibliography}[1]			
	{\ninerm
	 \baselineskip=11pt				
	 \begin{list}{\arabic{enumi}.}
	{\usecounter{enumi}\setlength{\parsep}{0pt}
	 \setlength{\leftmargin 17pt}{\rightmargin 0pt}	
	 \setlength{\itemsep}{0pt} \settowidth		
	{\labelwidth}{#1.}\sloppy}}{\end{list}}

\newcounter{itemlistc}
\newcounter{romanlistc}
\newcounter{alphlistc}
\newcounter{arabiclistc}

\newcommand{\fcaption}[1]{
        \refstepcounter{figure}
        \setbox\@tempboxa = \hbox{\eightrm Fig.~\thefigure. #1}
        \ifdim \wd\@tempboxa > 5in
           {\begin{center}
        \parbox{5in}{\eightrm \smalllineskip Fig.~\thefigure. #1 }
            \end{center}}
        \else
             {\begin{center}
             {\eightrm Fig.~\thefigure. #1}
              \end{center}}
        \fi}

\newcommand{\tcaption}[1]{
        \refstepcounter{table}
        \setbox\@tempboxa = \hbox{\eightrm Table~\thetable. #1}
        \ifdim \wd\@tempboxa > 5in
           {\begin{center}
        \parbox{5in}{\eightrm\smalllineskip Table~\thetable. #1 }
            \end{center}}
        \else
             {\begin{center}
             {\eightrm Table~\thetable. #1}
              \end{center}}
        \fi}

\def\@citex[#1]#2{\if@filesw\immediate\write\@auxout	
	{\string\citation{#2}}\fi			
\def\@citea{}\@cite{\@for\@citeb:=#2\do			
	{\@citea\def\@citea{,}\@ifundefined		
	{b@\@citeb}{{\bf ?}\@warning
	{Citation `\@citeb' on page \thepage \space undefined}}
	{\csname b@\@citeb\endcsname}}}{#1}}

\newif\if@cghi
\def\cite{\@cghitrue\@ifnextchar [{\@tempswatrue
	\@citex}{\@tempswafalse\@citex[]}}
\def\citelow{\@cghifalse\@ifnextchar [{\@tempswatrue
	\@citex}{\@tempswafalse\@citex[]}}
\def\@cite#1#2{{$\null^{#1}$\if@tempswa\typeout
	{IJCGA warning: optional citation argument
	ignored: `#2'} \fi}}

\def\pmb#1{\setbox0=\hbox{#1}
	\kern-.025em\copy0\kern-\wd0
	\kern.05em\copy0\kern-\wd0
	\kern-.025em\raise.0433em\box0}

\def\fnt#1#2{\footnotetext{\kern-.3em
	{$^{\mbox{\scriptsize #1}}$}{#2}}}

\def\fpage#1{\begingroup
\voffset=.3in
\thispagestyle{empty}\begin{table}[b]\centerline{\footnotesize #1}
	\end{table}\endgroup}

\def\runninghead#1#2{\pagestyle{myheadings}
\markboth{{\eightit{\quad #1}}\hfill}{\hfill{\eightit{#2\quad}}}}

\font\tenbf=cmbx10
\font\tenit=cmti10
\font\tenit=cmti10
\font\bfit=cmbxti10 at 10pt
 1
 1
 1

\font\ninerm=cmr9
\font\nineit=cmti9

\font\eightrm=cmr8
\font\eightit=cmti8

\def\qed{\hbox{${\vcenter{\vbox{                          
   \hrule height 0.4pt\hbox{\vrule width 0.4pt height 6pt
   \kern5pt\vrule width 0.4pt}\hrule height 0.4pt}}}$}}

\runninghead{On a New Method for Computing Trace Anomalies.}
{On a New Method for Computing Trace Anomalies.}

\begin{document}
\normalsize\textlineskip
{\thispagestyle{empty}
\setcounter{page}{1}

\renewcommand{\thefootnote}{\fnsymbol{footnote}} 

\copyrightheading{93-17}

\vspace*{0.88truein}

\fpage{1}
\centerline{\bf ON A NEW METHOD FOR COMPUTING}
\vspace*{0.035truein}
\centerline{\bf TRACE ANOMALIES}
\vspace{0.37truein}
\centerline{\footnotesize FIORENZO BASTIANELLI}
\vspace*{0.015truein}
\centerline{\footnotesize\it
Institute for Theoretical Physics,   Stockholm University}
\baselineskip=10pt
\centerline{\footnotesize\it  Box 6730,  S-113 85 Stockholm, Sweden}
\baselineskip=10pt
\centerline{\footnotesize\it  fiorenzo@vana.physto.se}
\vspace{.8truein}
\abstracts{\noindent
I describe a new method for computing trace anomalies in quantum field
theories which makes use of path-integrals for particles moving
in curved spaces. After presenting the main ideas of the method,
I discuss how it is connected to the first quantized approach of
particle theory and to heat kernel techniques.}{}{}

\vspace*{-3pt}\textlineskip
\section{Introduction}
\noindent
Systems coupled to gravity in a Weyl invariant way
(i.e. invariant under local rescalings of the metric)
have a vanishing trace of the energy-momentum tensor.
If one keeps the metric as a background,
then the system is invariant under the conformal group,
defined as the set of coordinate transformations
preserving the infinitesimal length up to a local scale factor.
However, anomalies may appear upon quantization. In particular,
performing a quantization which preserves general coordinate invariance
brings in an anomaly in the Weyl symmetry.
This is seen as a non-zero trace for the quantum energy-momentum tensor.
There are many ways to compute trace anomalies in
quantum field theory (Feynman diagrams,
heat kernel techniques, Fujikawa's approach, current algebras, etc.).
Recently, a new method has been proposed.$^{1,2}$
It employs the approach of Fujikawa$^3$ and can be described as follows.
{}From the path-integral one derives the Ward identities due
to the Weyl symmetry. Possible anomalies correspond to the
non-invariance of the functional measure and a Fujikawa jacobian is
obtained. This jacobian is suitably regulated by requiring general
coordinate invariance and \lq\lq consistency"
in the sense of Wess and Zumino. Then, its computation is
simplified enormously by noticing that it can be interpreted as a
trace over the Hilbert space of a \lq\lq fictitious" quantum mechanical
model having the given regulator as the hamiltonian. This trace is
computed using a path-integral representation. Thus,
one of the results in the analysis of refs. 1 and 2 is
a method for computing perturbatively path-integrals in curved spaces.
The method just described is a generalization to trace anomalies
of the ideas due to
Alvarez-Gaum\'e and Witten for computing chiral anomalies.$^4$
However, contrary to the chiral anomaly case, where a semiclassical
approximation is enough to get the complete
result valid in any space-time dimensions,
for trace anomalies one has to work harder: one must perform a
$(k+1)$-loop computation in quantum mechanics to obtain the (one-loop)
trace anomaly in $2k$ dimensions.
As already remarked in ref. 1, the \lq\lq fictitious" quantum mechanics
employed in these computations can be given a more
physical meaning by relating it to the first quantization of the
given quantum field theory. This point of view is interesting since
after the works of Bern and Kosower$^5$,
who derived string inspired Feynman rules to compute one-loop diagrams in
quantum field theory, and Strassler$^6$, who re-derived these rules
from first quantization, it seems important to look again
 at the first quantized picture. Understanding how the
perturbative expansion is organized in first quantization may suggest
several technical simplifications with respect to standard text-book rules
for computing loop diagrams. A simplification with respect to standard Feynman
rules is already seen here, in the computation of trace anomalies.
The first quantized approach is also instructive for
guessing  possible generalizations of the renormalization group equations
to string theory.$^7$

\textheight=7.8truein
\setcounter{footnote}{0}
\renewcommand{\thefootnote}{\alph{footnote}}

\section{The Method}
\noindent
A typical example of a classically Weyl invariant system is given
by the massless Klein-Gordon field $\phi$ coupled to gravity
\begin{equation}
     S[\phi,g] = \int d^D x {\sqrt g}\ {1\over 2} ( g^{\mu\nu}
     \partial_\mu \phi \partial_\nu \phi - \xi R \phi^2)
\end{equation}
where  $\xi= { D-2 \over 4(D-1)}$, $D$  denotes the space-time
dimensions and $R$ is the curvature scalar.
To quantize the Klein-Gordon field, one may
consider the euclidean path-integral
\begin{equation}
      Z[g] = {\rm e}^{-W[g]} = \int {\cal D} \phi \ {\rm e}^{-S[\phi,g]}.
\end{equation}
One derives the anomalous Ward identities for the Weyl symmetry
by  performing a dummy change of integration variables
$\phi \rightarrow \phi' =\phi +\delta \phi $, where $\delta \phi=
{2-D\over 2} \sigma \phi$ is an  infinitesimal Weyl transformation
with parameter $\sigma$,
and  using the invariance of $S[\phi,g]$.
If the Fujikawa variables $\tilde \phi = g^{1\over 4} \phi$
are used as integration variables
to avoid gravitational anomalies, one obtains
the following Ward identity for the effective action $W[g]$
\begin{equation}
     \delta W[g]
     =\int d^D x {\sqrt g} \sigma \langle \Theta \rangle
     = - {\rm \lq\lq Tr ( }  \sigma{\rm )"}
     = -  \lim_{\beta \to 0} {\rm Tr} \bigl ( \sigma
     {\rm e}^{ \beta(\nabla^2 +\xi R)}\bigr )
\end{equation}
where $\Theta$ denotes the trace of the energy-momentum tensor and where
the last step describes the regularization of the infinite dimensional
functional trace, $\beta$ being the regulating parameter
and $ (\nabla^2 +\xi R)$ the consistent and general coordinate invariant
regulator.
The trace on the right hand side of this chain of identities
can be evaluated using a path-integral representation
\begin{equation}
      {\rm Tr} \bigl ( \sigma {\rm e}^{-\beta H } \bigr ) =
      \int_{PBC} ({\cal D}q)\  \sigma {\rm e}^{ - S[q]}
\end{equation}
where I have identified $H= -(\nabla^2 +\xi R)$ as the quantum hamiltonian
for a particle propagating in curved space for an euclidean time $\beta $
and with periodic boundary conditions ($PBC$).
In this formula the action for the particle is given by
\begin{equation}
        S[q] = \int_{0}^{\beta} dt\ \bigl ({1\over 2}
        g_{\mu\nu}(q) \dot q^\mu \dot q^\nu + V(q) \bigr )
\end{equation}
where $V(q)$ is a scalar potential to be specified shortly, and the
functional measure is the general covariant one
$({\cal D}q) = \prod_{t} {\sqrt{ g(q(t))}} d^D q(t)$.
As it stands, the path-integral is still a formal
expression as one must tell how to compute it.
It is known that different prescriptions for discretizing the path-integral
correspond in canonical quantization to different orderings of
coordinates and momenta in the quantum hamiltonian.
Therefore, one has to identify correctly the hamiltonian associated
with the chosen discretization.
The strategy followed in refs. 1 and 2
was to compute the path-integral perturbatively in $\beta$ (used
as a loop-counting parameter) and expanding the quantum
fluctuations of the coordinates $q^\mu(t)$
around the classical path $q^\mu_{cl}(t)$ in a sine series
\begin{equation}
       q^\mu(t) = q^\mu_{cl}(t) + q^\mu_{qu}(t),\ \ \ \
       q^\mu_{qu}(t) = \sum_{n=1}^{M} q^\mu_n \sin {\pi  n t\over \beta}.
\end{equation}
The truncation of the mode expansion at a fixed mode $M$ acts as
a discretization while the measure $({\cal D}q)$ is expressed in terms of
the Fourier coefficients $q^\mu_n$. The continuum limit is achieved
by letting $M\to \infty$. Also,
ghost fields were used to exponentiate the non-trivial part of the
measure, so that standard perturbation theory could be applied
(by getting propagators form the quadratic part of the action and vertices
from the rest). Using Riemann normal coordinate to simplify computations,
it can be checked that the transition amplitudes
given by this path-integral satisfies an
euclidean Schroedinger equation (i.e. a heat equation)
with
hamiltonian $H=-(\nabla^2 +\xi R)$ once the scalar potential
$V= {1\over 2} ({1\over 4}-\xi) R$ is used in (5).
Having made explicit the relation between the canonical
and lagrangian quantization for the particle in a curved background,
one can proceed and compute the trace anomalies in (3).
Since the path-integral measure expressed in terms
of the Fourier coefficients $q^\mu_n$
carries a normalization factor
$A= {(2 \pi\beta)}^{-{D\over2}}$,
one has to compute the (${D\over2}+1$)-loop \lq\lq vacuum" graphs to cancel
the $\beta$-dependence of the measure and obtain a finite result
(in the loop expansion propagators
carry a factor $\beta$ and vertices a factor $\beta^{-1}$).
This way the trace anomalies were computed in two and four dimensions,
and for spin $1\over 2$ and $1$ as well,
reproducing the known values.

\section{Connection to First Quantization}
\noindent
I have described  a method for computing trace anomalies
in quantum field theories which makes use of path-integrals
for non-relativistic particles moving in curved spaces.
In such a pragmatic approach,
 one considers the auxiliary quantum mechanical system,
the particle in a curved background,
to be of a fictitious, non-physical nature,
as it is usually regarded in the Schwinger-De Witt
heat kernel approach.$^8$
At a closer inspection, however, one can reinterpret the method
as being the evaluation of the Weyl anomaly  using  the first
quantized  description of a scalar field coupled to gravity.
This interpretation can be extended also to the  Schwinger-De Witt
method for calculating the Klein-Gordon propagator in curved spaces.$^8$
In that method, the Klein-Gordon  propagator is obtained by integrating
over the proper time a kernel which satisfies the heat equation.
Such an equation is solved by an ansatz depending on certain
coefficients $a_{k}$, called sometimes Seeley-De Witt coefficients.
The solution is then used to construct the one-loop effective action $W[g]$.
Computing the Weyl variation of the effective action,
 one obtains precisely the Seeley-De Witt
coefficient  $a_{k}$ as the trace anomaly in  $2k$
space-time dimensions. The method described in the previous section gives
a way of computing directly the Seeley-De Witt coefficients by
using the path-integral representation of the solution to the heat equation.
As for the heat kernel method used for constructing the
effective action, it can be derived from first quantization by computing
\begin{equation}
W[g] = \int_{PBC} {({\cal D}e) ({\cal D}q)\over {{\rm vol}\  {\rm Diff}}}
\ {\rm e}^{ - S[e,q]}
\end{equation}
where $S[e,q]$ is the action for a scalar particle
obtained by covariantizing (5) on the world-line with the help
of the einbein $e$. After gauge fixing, the integration over the einbeins
reproduces the integration over the proper time.
These arguments are enough to describe here
the relation between the given method for calculating trace anomalies,
the heat-kernel techniques of Schwinger-De Witt and the
first quantized approach to particle theory.
Further details will be  presented in a future publication.$^9$

\nonumsection{Acknowledgements}
\noindent
I wish to thank P. van Nieuwenhuizen for his collaboration on this project.

\nonumsection{References}

\end{document}